\documentclass[%
 reprint,
superscriptaddress,
 amsmath,amssymb,
 aps,prd
]{revtex4-2}

\usepackage{graphicx}
\usepackage{dcolumn}
\usepackage{bm}
\begin{document}


\title{The overtone level spacing of a black hole quasinormal frequencies: a fingerprint of a local $SL(2,\mathbb{R})$ symmetry 
} 

\author{Bernard Raffaelli}
\email{bernard.raffaelli@u-bourgogne.fr}
\affiliation{Institut de Math\'{e}matiques de Bourgogne, UMR 5584 CNRS, Universit\'{e} Bourgogne Franche-Comt\'{e}, F-2100 Dijon, France}

\date{\today}

\begin{abstract}
The imaginary part of the quasinormal frequencies spectrum for a static and spherically symmetric black hole is analytically known to be equally spaced, both for the highly damped and the weakly damped families of quasinormal modes. Some interesting attempts have been made in the last twenty years to understand in simple ways this level spacing for the only case of highly damped quasinormal frequencies. Here, we show that the overtone level spacing, for both the highly damped and weakly damped families of quasinormal modes, can simply be understood as a fingerprint of a hidden local $SL(2,\mathbb{R})$ symmetry, near different regions of the black hole spacetime, i.e. the near-horizon and the near-photon sphere regions.
\end{abstract}

\maketitle


\section{\label{sec:Intro}Introduction}

In the context of small perturbations of a black hole (BH) spacetime, the quasinormal modes (QNM), which are damped modes of perturbation which propagate towards spatial infinity, are playing a fundamental role both in our experimental and theoretical understanding of gravity and of BH in particular. 
The QNM are related to complex quasinormal frequencies (QNF) at which they occur. The real part of the QNF corresponds to the frequency of oscillation and the imaginary part to the damping rate of the QNM.
Analytically, for example in the case of static and spherically symmetric (asymptotically flat) BH perturbed by a massless scalar field, two families of QNM are known 
\begin{itemize}
\item the \emph{highly damped QNM} with the associated QNF (at first order in $n$)
\begin{equation}\label{HighlyDampedQNF}
    \omega_n = \frac{\kappa}{2\pi}\ln\left[1+2\cos\left(\frac{\pi(qD-2)}{2}\right)\right] - i\kappa \left(n+\frac{1}{2}\right)
\end{equation}
where $\kappa$ is the surface gravity of the BH, $D=d-2$ for $d$-dimensional spacetimes and $q$ is a purely real power law exponent that characterizes the BH metric near its singularity \cite{DasShankaranarayananDas2005,GoshShankaranarayananDas2006}, 
\item and the \emph{weakly damped QNM} with the associated QNF (at first order in $\ell$ and $n$)
\begin{equation}\label{WeaklyDampedQNF}
\omega_{\ell,n}=\Omega_c\left(\ell+\frac{1}{2}\right)-i|\Lambda_c|\left(n+\frac{1}{2}\right)    
\end{equation}
where $\Omega_c$ is the angular velocity of massless particles on unstable circular orbits around the BH, and $|\Lambda_c|$ is the Lyapunov exponent associated with these unstable circular orbits \cite{CardosoMirandaBertiWitekZanchin2009,DecaniniFolacciRaffaelli2010}.
\end{itemize}
The imaginary part of the QNF, in both cases, is interesting. It contains the overtone number $n$ and is equally spaced. Some remarks about its amplitude will also be given in this paper. Then, it would be natural to ask whether the seemingly simple behavior of the imaginary part of the QNF in both cases may, or may not, hide a ``universal'' feature of BH spacetimes. For static and spherically symmetric spacetimes, such an imaginary part for the BH QNF, in the only case of the highly damped QNM, has been discussed in \cite{Padmanabhan2004}, where it has been obtained as poles of the scattering amplitude in the Born approximation, and in \cite{BaticKelkarNowakowski2011} where it has been extracted via a Coulomb-like phase shift.\\
In this paper, we would like to show that such a behavior for the imaginary part of a BH QNF is in fact more ``universal'' and can simply be understood, in both the highly and weakly damped regimes, as a fingerprint of a hidden local $SL(2,\mathbb{R})$ symmetry. We will focus on static and spherically symmetric BH endowed with a photon sphere (also known as ``lightrings'', or ``photon rings''). The strategy is made simple as long as one can show that certain particular regions of a BH spacetime admit a Rindler metric as an approximation. Indeed, on the one hand, the resonant scattering of a massless scalar field in a Rindler spacetime is shown to be deeply linked to the resonant scattering problem of a scalar field by an inverted harmonic oscillator (IHO) potential of the form $V(x)=-(\alpha^2/4) x^2$, where $\alpha$ is the potential curvature, or potential strength. On the other hand, the resonant scattering problem by an IHO potential can be algebraically solved using the $SL(2,\mathbb{R})$ algebra for which the representations and the eigenvalues spectrum are discrete and indexed by a non negative integer. The general claim of this paper follows: in any BH geometry, if there exist some regions where the BH metric can be locally approximated as a Rindler one, in some set of coordinates, then the underlying algebra is $SL(2,\mathbb{R})$, and the imaginary part of the BH QNF will be indexed by a non negative integer, i.e. the overtone number, and it will be equally spaced. The amplitude of the overtone level spacing is shown to be related to the IHO potential curvature.\\ 

\noindent The paper is organized as follows. In section \ref{sec:IHO}, we give an $SL(2,\mathbb{R})$ approach to the resonant scattering problem of a scalar field by an IHO potential. This allows to highlight that the eigenvalues spectrum of the corresponding Hamiltonian is discrete, purely imaginary and indexed by a non negative integer. In section \ref{sec:Rindler}, we focus first on the motions of massless particles in Rindler spacetime, and then on the resonant scattering problem of a massless scalar field in such a spacetime. We show that massless dynamics in Rindler spacetime is deeply linked to the problem treated in section \ref{sec:IHO}, and we apply the results we have obtained to the Rindler case. Finally, in section \ref{sec:SSSBH}, we apply the $SL(2,\mathbb{R})$ algebra to the case of static and spherically symmetric BH endowed with a photon sphere, and show that the imaginary part of the BH QNF in both the highly an weakly damped regimes are equally spaced. In sections \ref{sec:Rks} and \ref{sec:Cl} we discuss the amplitude of the overtone level spacing and gives somes conclusions about the results obtained in this paper.\\


%
%

\section{\label{sec:IHO}An $SL(2,\mathbb{R})$ approach to the resonant scattering by an inverted harmonic oscillator potential}
With a particular choice of ``position'' operator $X$ and conjugate ``linear momentum'' operator $P$, acting both on a same Hilbert space, the Hamiltonian of a particle in an IHO potential can always be written as
\begin{equation}\label{Hamiltonian}
H=P^2 - \frac{\alpha^2}{4}X^2,
\end{equation}
where $\alpha$ is the curvature of the IHO potential, and the operators, which satisfy $[X,P]=i\mathbb{I}$, are defined, as usual, as $P=-i\frac{d}{dx}$ and $X=x$, in an ``$x$-representation''. In a system of units such that $h=c=1$, it should be noted that if one wants $H$ to have dimension $L^{-1}$, $P$ and $X$ have to be defined here with dimensions $L^{-1/2}$ and $L^{1/2}$ respectively. The IHO potential curvature $\alpha$ has dimension $L^{-1}$.\\

\noindent From the point of view of a scalar field theory, the Hamiltonian gives the following equation for its eigenstates $\psi$ in the $x$-representation
\begin{equation}\label{ScalarFieldEq}
H \psi(x)=\left(-\frac{d^2}{d x^2}-\frac{\alpha^2}{4}x^2\right)\psi(x)=h \psi(x).
\end{equation}
Here $h$ is the Hamiltonian eigenvalue associated with the eigenstate $\psi$. This equation is a Weber differential equation whose solutions are related to parabolic cylinder functions \cite{GradshteynRyzhikZwillingerMoll2014}. They describe the scattering states of the scalar field by an IHO potential. Amongst the scattering states, the system presents a discrete set of eigenstates, i.e. the resonant modes, or QNM, which can be obtained, for example, by looking at the poles of the corresponding $S$-matrix in the complex energy plane.\\

\noindent An expression of $H$ in the form $(\ref{Hamiltonian})$ (or $(\ref{ScalarFieldEq})$) plays an ``universal'' role for the resonant scattering problem of a scalar field by the top of any potential barrier. Indeed, in such case, the QNM are associated with energies close to the top of the potential. For such energies, one can always approximate any potential barrier close to its maximum by a second order Taylor series expansion which locally gives an inversed parabola. With the right choice of $X$ and $P$ operators, the corresponding Hamiltonian $H$ can then always be reduced to the form $(\ref{Hamiltonian})$ (or $(\ref{ScalarFieldEq})$). This form of $H$ is not only ``universal'' for a problem of resonant scattering by a potential barrier, but it also permits to reveal a hidden (dynamical) conformal symmetry near the top of the potential through an $SO(2,1) \sim SL(2,\mathbb{R})$ algebra which allows to obtain the associated set of QNM from the algebra representations and, especially for the aim of this paper, to obtain the overtone level spacing in the imaginary part of the quasinormal frequencies. We briefly recall below the outline of one of the possible proofs that mainly lies on the introduction of ``creation and annihilation operators'' analogues in the factorization of $H$, which allows to highlight the underlying $SL(2,\mathbb{R})$ algebra (see \cite{Raffaelli2022} for more details).\\

We have seen that the time-independent, one-dimensional Hamiltonian of a quantum particle scattered by an IHO potential reads in the $x$-representation
\begin{equation}\label{DimensionlessHamitonian}
H=-\frac{d^2}{dx^2} - \frac{\alpha^2}{4}x^2.
\end{equation}

As one usually proceeds in the context of the well-known harmonic oscillator problem, let us begin by factorizing $H$. From the operators $P=-i\frac{d}{dx}$ and $X=x$ (mentioned in the previous section), one can introduce ``creation and annihilation operators'' analogues
\begin{equation}\label{CreationAnnihilationOperators}
U_{\pm}=\pm \frac{1}{\sqrt{\alpha}}P+\frac{\sqrt{\alpha}}{2}X=\mp i\frac{1}{\sqrt{\alpha}}\frac{d}{dx}+\frac{\sqrt{\alpha}}{2}x,
\end{equation} 
with $(U_{\pm})^{\dagger} = U_{\pm} \neq U_{\mp}$ (with respect to the usual scalar product) and $[U_{-},U_{+}]=i\mathbb{I}$. Let us note that in our system of units, $U_{\pm}$ are dimensionless operators. The Hamiltonian $H$ can then be written in terms of $U_{\pm}$ as
\begin{equation}\label{H_Upm}
H=P^2-\frac{\alpha^2}{4}X^2=-\frac{\alpha}{2}\left(U_{+}U_{-}+U_{-}U_{+}\right).
\end{equation} 
It is also possible to introduce the dilation operator $D$ and the operator $S$, which is associated with special conformal transformations generator $K=X^2$, as follows
\begin{eqnarray}\label{DS_Upm}
&D&=\frac{1}{2}\left(PX+XP\right)=\frac{1}{2}\left(U_{+}^2-U_{-}^2\right)=-i\left(x\frac{d}{dx} + \frac{1}{2}\right),\nonumber\\
&S&=P^2+\frac{\alpha^2}{4}X^2=\frac{\alpha}{2}\left(U_{+}^2+U_{-}^2\right)=-\frac{d^2}{dx^2}+\frac{\alpha}{4}x^2.
\end{eqnarray}
Up to dimensional factors, it is important to note that the expressions of $H$ and $D$ are switched whether one is looking at their expressions in terms of $X$ and $P$ operators, or in terms of $U_{\pm}$ operators. In other words, for what concerns the IHO problem, $H$ can be seen as a dilation process or as a scattering process by an IHO potential, depending on the choice of ``coordinates'' ($x$ or $u_{\pm}$) and the associated conjugate momenta ($p$ or $u_{\mp}$). It is interesting to note that the operator $S$ remains unchanged in both ``coordinates'' systems, as a sum of squared operators, and that it would play the role of the Hamiltonian in the usual harmonic oscillator problem. It becomes then quite trivial to write the expressions of $H$, $D$ and $S$ as differential operators in ``$u_{\pm}$''-representations, from their expressions in the $x$-representation (see \cite{SubramanyanHedgeVishveshwaraBradlyn2019} and references therein, for a nice exhaustive analysis of the physics of the IHO). For example, $H$ in the $u_{\pm}$-representation reads simply

\begin{equation}\label{H_Diff_Upm}
H = \mp i \alpha\left(u_{\pm}\frac{d}{du_{\pm}}+\frac{1}{2}\right).
\end{equation} 

In order to simplify the underlying algebra, one can finally define the following three (dimensionless) operators
\begin{subequations}\label{JOperators}
\begin{eqnarray}
J_1&=&\frac{1}{2}D=-\frac{i}{2}\left(x\frac{d}{dx}+\frac{1}{2}\right),\\ 
J_2&=&-\frac{i}{2\alpha}S=\frac{i}{2\alpha}\left(\frac{d^2}{dx^2}-\frac{\alpha^2}{4}x^2\right),\\ 
J_3&=&\frac{i}{2\alpha}H=-\frac{i}{2\alpha}\left(\frac{d^2}{dx^2}+\frac{\alpha^2}{4}x^2\right).
\end{eqnarray}
\end{subequations}
Those operators satisfy the commutation relations of an $SO(2,1)$ algebra
\begin{equation}
[J_1,J_2]=-iJ_3; \quad [J_2,J_3]=iJ_1; \quad [J_3,J_1]=iJ_2.
\end{equation}
Finally, from $J_1$ and $J_2$, one can also introduce ``ladder operators''. For example
\begin{equation}\label{J+-Operators}
J_{\pm}=\pm iJ_1-J_2=\frac{i}{2}U_{\pm}^2,
\end{equation}
which, together with $J_3$, satisfy an $SL(2,\mathbb{R})$ algebra
\begin{equation}\label{LadderOp_CommutationRelation}
[J_{+},J_{-}]=-2J_3; \quad [J_3,J_{\pm}]=\pm J_{\pm}.
\end{equation}
It should be noted first that one could have obviously constructed $J_3$ and $J_{\pm}$, which satisfy commutation relations $(\ref{LadderOp_CommutationRelation})$, more directly from $H$ and $U_{\pm}^2$, i.e. without any references to the operators $D$ and $S$. Finally, let us also note that that the ladder operators $J_{\pm}$ are not self-adjoint conjugate to each other. In our example here, one has instead $\left(J_\pm\right)^\dagger=-J_\pm$, i.e. anti-self-adjoint operators.\\

Once the algebra is found, the computation of the eigenstates and eigenvalues of $H$ follows. Indeed, as usually done in the harmonic oscillator case, one can begin by looking for the ground states of the ladder operators $J_\pm$. Then, by an application of $(J_\pm)^n$ to the associated ground state (for every non negative integer $n$), it is an easy task to construct the set of eigenstates which belong to the associated rigged Hilbert space, that allows, in a few words, states to be defined as distributions (see \cite{Bohm1981,CivitareseGadella2004} for more details about the rigged Hilbert space formalism). The interested reader may found a detailed construction of the set of eigenstates and eigenvalues of $(J_\pm)^n$ and $H$ in \cite{Raffaelli2022}, in the context of a hidden $SL(2,\mathbb{R})$ symmetry on a BH photon sphere. Finally, one shows that the associated spectrum of $H$ is discrete, purely imaginary, and indexed by an integer. The eigenvalues $h_n^{(\pm)}$ of $H$, for every non negative integer $n$, can finally be written as
\begin{equation}\label{HEigenvalues}
h_n^{(\pm)} = \mp i \alpha \left(n+\frac{1}{2}\right).
\end{equation}
These discrete and purely imaginary eigenvalues are shown to be related to the frequencies of the resonant modes in the resonant scattering problem of a scalar field by an IHO potential \cite{SubramanyanHedgeVishveshwaraBradlyn2019, Raffaelli2022}. The non negative integer index $n$ will be the key to understand the overtone level spacing of a BH quasinormal frequencies. It should be noted that the curvature $\alpha$ of the IHO potential appears to be the amplitude of the eigenvalues spacing of $H$.\\

In the following, we will first show that the resonant scattering problem of a massless scalar field in a Rindler spacetime can precisely be reduced to the study of a resonant scattering problem by an IHO potential with the corresponding Hamiltonian $H$. From the above, this obviously allows its algebraic treatment through the $SL(2,\mathbb{R})$ algebra with all the related results. By a natural extension, it should be noted that every regions of a black hole spacetime that locally admit the Rindler metric as an approximation will be places where a similar treatment can be applied. In particular, we will show in the following sections that the horizon (which is the best known case) but also the photon sphere are both such regions.

\section{\label{sec:Rindler}The Rindler spacetime case}
\subsection{\label{subsec:MotionInRindler} Massless particles in Rindler spacetime}
We consider the line element of a $(1+1)$-dimensional Rindler spacetime, in Rindler's coordinates $(t,x)$:
\begin{equation}\label{RindlerMetric}
ds^2=a^2x^2 dt^2-dx^2.
\end{equation}
It is known to describe a line element seen by an uniformly accelerated observer (with a constant acceleration $a$), in coordinates $(t,x)$. The study can be restricted, for example, to the right Rindler wedge, for which $t\in]-\infty,+\infty[$ and $x\in[0,+\infty[$. In the following, we will be focusing on the motions of massless particles in those coordinates. With this aim in mind, we introduce an affine parameter $\lambda$ to describe the geodesics, and we decide to use a (quadratic) Lagrangian approach which is a minimalist but efficient way to highlight our point. Let us consider the following quadratic lagrangian:
\begin{equation}\label{QuadraticLagragian}
\mathcal{L}=\frac{1}{2}\frac{ds^2}{d\lambda^2}=\frac{1}{2}(a^2x^2\dot{t}^2-\dot{x}^2)
\end{equation}
where $\dot{t}=dt/d\lambda$ and $\dot{x}=dx/d\lambda$. In units such that $h=c=1$, here $x$ and $t$ have dimension $L$, $\lambda$ has dimension $L^2$, while $a$ has dimension $[x]^{-1}=L^{-1}$.\\

\noindent From Euler-Lagrange equations and spacetime symmetries, the only integral of motion (equivalently associated with the Killing vector $\partial/\partial t$) is here the ``energy'' $E$ defined by
\begin{equation}\label{E}
E=\frac{\partial \mathcal{L}}{\partial \dot{t}}=a^2x^2\dot{t}.
\end{equation}
The ``linear momentum'' is defined as
\begin{equation}\label{p}
p=\frac{\partial \mathcal{L}}{\partial \dot{x}}=-\dot{x}.
\end{equation}
Inserting $(\ref{E})$ into the line element $(\ref{RindlerMetric})$ for a massless particle, i.e. $ds^2=0$, one finds
\begin{equation}\label{MasslessMotion1}
\dot{x}^2 - \frac{E^2}{a^2x^2}=0
\end{equation}
or 
\begin{equation}\label{MasslessMotion2}
\left(\frac{dx}{dt}\right)^2 - a^2x^2=0.
\end{equation}

This equation, whose solution grows exponentially with an associated Lyapunov exponent $\Lambda=|a|$ (see \cite{DaluiMajhi2020} and references therein), describes the trajectory of a massless particle in Rindler's coordinates $(t,x)$, i.e. seen by an uniformly accelerated observer. Equation $(\ref{MasslessMotion2})$ clearly shows that this motion, in $(t,x)$ coordinates, is actually a motion in an IHO potential. This equation can then also be interpreted as describing an instability near the (hyperbolic) point $x=0$ (the top of the IHO potential barrier) in the phase space $(dx/dt,x)$. In other words, the Rindler horizon corresponds to an unstable equilibrium point for massless particles, in the phase space $(dx/dt,x)$ of an accelerated observer. Let us note that for massless geodesics ($ds^2=0$), dividing $(\ref{RindlerMetric})$ by $dt^2$ would have given directly $(\ref{MasslessMotion2})$.\\

It should be noted that there exist a simple dispersion relation between the energy $E$ and the linear momentum $p$ of the particle, obtained by inserting $(\ref{p})$ in $(\ref{MasslessMotion1})$
\begin{equation}\label{RindlerDispersionRelation}
E=\pm a x p.
\end{equation}
This dispersion relation is characteristic of a particle in a IHO potential. Indeed, with a change of variables of the form
\begin{equation}
\tilde{x}=x-\frac{2}{a}p \quad \textrm{and}\qquad \tilde{p}=\frac{a}{2}x + p,
\end{equation}
which is equivalent to $(\ref{CreationAnnihilationOperators})$, one can rewrite $(\ref{RindlerDispersionRelation})$ as
\begin{equation}
E = \pm \left(\tilde{p}^2-\frac{a^2}{4}\tilde{x}^2\right).
\end{equation}
where the IHO potential appears explicitly.
 
\subsection{\label{subsec:ScatteringRindler} Scattering of a massless scalar field in a Rindler spacetime}
The Klein Gordon equation for a massless scalar field $\Phi(x^\mu)$ in a spacetime with metric $g_{\mu\nu}$ reads
\begin{equation}\label{KleinGordon}
\Box \Phi = g^{\mu\nu} \nabla_\mu \nabla_\nu \Phi = \frac{1}{\sqrt{-g}} \partial_\mu \left(\sqrt{-g} g^{\mu\nu}\partial_\nu \Phi\right)=0.
\end{equation}
In a Rindler spacetime with metric $(\ref{RindlerMetric})$, the Klein Gordon equation becomes
\begin{equation}\label{KleinGordon_Rindler}
\left[-\partial_t^2 + a^2x\partial_x\left(x\partial_x\right)\right]\Phi(t,x)=0.
\end{equation}
Let us assume a stationary scalar field with harmonic time dependence of the form $\Phi(t,x)=\phi(x)e^{-i\omega t}$. The Klein-Gordon equation becomes
\begin{equation}
\left[(-i\omega)^2 - ax\partial_x\left(x\partial_x\right)\right]\Phi(t,x)=0,
\end{equation}
which can be easily factorized in the form
\begin{equation}\label{KG_factorized}
\left[(-i\omega)+ax\partial_x\right]\left[(-i\omega)-ax\partial_x\right]\Phi(t,x)=0.
\end{equation}
It should be noted that a massless scalar plane wave, e.g. $\Phi(t,x)=\Phi_0 e^{-i(\omega t \mp kx)}$, solution of each term of $(\ref{KG_factorized})$ gives a dispersion relation
\begin{equation}
\omega = \pm a x k
\end{equation}
which is equivalent to the relation $(\ref{RindlerDispersionRelation})$.\\

\noindent In order to establish explicitly the link between $(\ref{KG_factorized})$ and $(\ref{H_Diff_Upm})$, we will have to introduce a ``trick'' that will highlight an important aspect of thermality comparing the Rindler and the IHO case. Let us begin to rewrite $(\ref{KG_factorized})$ as
\begin{equation}\label{KG_factorized_2}
\left[-i\omega-\frac{a}{2}+\frac{a}{2}+ax\partial_x\right]\left[-i\omega+\frac{a}{2}-\frac{a}{2}-ax\partial_x\right]\Phi(t,x)=0.
\end{equation}
Most of the solutions $\Phi$ \cite{MorettiPinamonti2003} can be reduce to a linear combination of $\tilde{\Phi}^{(\pm)}$, such as
\begin{equation}\label{Weyl_like_eqs1}
\left\{
\begin{array}{ll}
\left(ax\partial_x + \dfrac{a}{2}\right)\tilde{\Phi}^{(+)}=\left(i\omega+\dfrac{a}{2}\right)\tilde{\Phi}^{(+)}\\
-\left(ax\partial_x + \dfrac{a}{2}\right)\tilde{\Phi}^{(-)}=\left(i\omega-\dfrac{a}{2}\right)\tilde{\Phi}^{(-)}.
\end{array}
\right.
\end{equation}
In other words, $\tilde{\Phi}^{(\pm)}$ both satify
\begin{equation}\label{Weyl_like_eqs2}
\pm ia\left(x\partial_x +\frac{1}{2}\right)\tilde{\Phi}^{(\pm)} = - \left(\omega \mp \frac{ia}{2}\right)\tilde{\Phi}^{(\pm)},
\end{equation}
which reads
\begin{equation}\label{H_Rindler_IHO}
H \tilde{\Phi}^{(\pm)} = \left(\omega \mp \frac{ia}{2}\right)\tilde{\Phi}^{(\pm)}
\end{equation}
where $H$ is the Hamiltonian of an IHO written in the form $(\ref{H_Diff_Upm})$, i.e. as a dilation operator. In other words, the equation of motion of a massless scalar field, of frequency $\omega$, seen by an uniformly accelerated observer (with acceleration $a$) in Rindler coordinates $(t,x)$, can simply be expressed in term of an IHO Hamiltonian. It should be noted that the naive ``trick'' used in $(\ref{KG_factorized_2})$ that leads to $(\ref{H_Rindler_IHO})$, could be understood as being related to the thermal behavior of the number of quantized particles in the Unruh effect. Indeed, on the one hand, the factorization of the massless Klein Gordon equation for a scalar field $\Phi$ in a $(1+1)$-dimensional Rindler spacetime with metric $(\ref{RindlerMetric})$ gives a system of two massless scalar field $\tilde{\Phi}^{(\pm)}$ equations, the same way the usual $(1+1)$-dimensional massless scalar wave equation can be factorized into a system of two $(1+1)$-dimensional uncoupled Weyl-like equations. On the other hand, it is well-known that, for a scalar field, the thermal behavior of the number of quantized particles seen by an accelerated observer is given by a Bose-Einstein factor $(e^{2\pi\omega/T}-1)^{-1}$ with Unruh temperature $T$, whereas for a spin-$1/2$ field the thermal behavior is given by a Fermi-Dirac factor $(e^{2\pi\omega/T}+1)^{-1}$. From this perspective, the ``trick'' in $(\ref{KG_factorized_2})$ formally corresponds to the passage from one thermal behavior, that originates from equation $(\ref{KleinGordon_Rindler})$, to the other, that originates from equations $(\ref{Weyl_like_eqs1})$, with the change $\omega \to \omega\mp ia/2$ in the computation of the related Unruh effect \citep{AlsingMilonni2004}.\\

\noindent We have explicitly shown in this section that the scattering by an IHO potential is actually hidden in the scattering problem of a massless scalar field in Rindler spacetime, in coordinates $(t,x)$. The resonant scattering in Rindler spacetime is then simply given by the $SL(2,\mathbb{R})$ algebraic approach presented in section \ref{sec:IHO}. Indeed, by combining $(\ref{HEigenvalues})$ and $(\ref{H_Rindler_IHO})$ gives a set of purely imaginary frequencies, for every non negative integer $n$
\begin{equation}\label{Rindler_QNF}
\omega_{n}^{(\pm)}=\mp i a n.
\end{equation}
As anticipated previously, it should be noted that the discrete eigenvalues spectrum of the IHO Hamiltonian for the associated resonant modes can be understood as being at the origin of the regular level spacing in the QNF. This level spacing appears clearly here, in the case of resonant scattering in Rindler spacetime. Let us note that the amplitude of the level spacing is nothing less than the IHO potential curvature, which is interpreted as a constant acceleration in the Rindler case. This will appear in a very similar manner in the QNF spectrum for BH, as long as the BH metric admits close to certain regions of spacetime, the Rindler metric as an approximation.

\section{\label{sec:SSSBH}The static and spherically symmetric black hole case}

In the Schwarzschild coordinates $(t,r,\theta,\phi)$, we consider a static spherically symmetric four-dimensional spacetime with metric
\begin{equation}\label{metric_BH}
ds^2=f(r)dt^2-\frac{dr^2}{f(r)}-r^2d\sigma^{2}.
\end{equation}
As usual, $d\sigma^{2}=d\theta^2 + \sin^2 \theta d\varphi^2$ is the line element on the unit 2-sphere $S^{2}$, with $\theta \in [0,\pi]$ and $\varphi \in [0,2\pi]$.\\
Moreover, let us consider the BH exterior with $r \in ]r_h,+\infty[$, where $r=r_h$ is a simple root of $f(r)$ and defines the location of the BH event horizon. Let us assume also that the background geometry is asymptotically flat and the tortoise coordinate $r_\ast = r_\ast(r)$, defined as $dr_\ast/dr = 1/f(r)$, is a bijection from $]r_h,+\infty[$ to $] -\infty, +\infty[$.\\

Without loss of generality, we will consider motions on the equatorial plane $\theta=\pi/2$. A free-falling massless particle moves along null geodesics according to
\begin{equation}\label{NullGeodesics}
-f(r)\dot{t}^2+\frac{1}{f(r)}\dot{r}^2+r^2\dot{\varphi}^2=0,
\end{equation}
where, as done earlier in this paper,  $\dot{t}=dt/d\lambda$, $\dot{r}=dr/d\lambda$, $\dot{\varphi}=d\varphi/d\lambda$ and $\lambda$ is an affine parameter which describes null geodesics.\\ 

\noindent From symmetries of the BH spacetime, one can define integrals of motion, i.e. energy $E$ and angular momentum $L$ of the massless particle, associated respectively with the Killing vectors $\partial/\partial t$ and $\partial/\partial \varphi$:
\begin{equation}\label{EL}
f(r)\dot{t}=E\quad ; \quad r^2\dot{\varphi}=L.
\end{equation}
The equation of motion is easily deduced from (\ref{NullGeodesics}) and reads
\begin{equation}\label{EqMotion}
\dot{r}^2+V_{\textrm{eff}}(r)=E^2,
\end{equation}
where the effective potential $V_{\textrm{eff}}$ is defined as
\begin{equation}\label{EffPotential}
V_{\textrm{eff}}(r)=\frac{L^2}{r^2}f(r).
\end{equation}

\noindent A photon sphere (or ``lightrings'', or ``photon rings'') located at $r=r_c$ corresponds to a local maximum of $V_{\textrm{eff}}(r)$ at $r=r_c$ such as
\begin{subequations}
\begin{eqnarray}
\left.\frac{d}{dr}V_{\textrm{eff}}(r)\right|_{r_c}=0 &\Leftrightarrow& \frac{2}{r_c}f_c=f'_c, \label{rc}\\
\left.\frac{d^2}{dr^2}V_{\textrm{eff}}(r)\right|_{r_c}<0 &\Leftrightarrow& f''_c - \frac{2}{r_c^2}f_c<0.
\end{eqnarray}
\end{subequations}
The subscript ``$c$'' means, above and in the following, that the quantity considered is evaluated at $r=r_c$, and the superscripts `` $'$ '' and `` $''$ '' respectively mean the first and second derivatives with respect to $r$.\\

A massless particle reaches the photon sphere when the turning point of its motion satisfies $E^2 = V_\textrm{eff,c} $, i.e. $L/E = r_c/\sqrt{f_c}$, where $r_c/\sqrt{f_c}=b_c$ is the critical impact parameter for massless particles to reach tangentially the photon sphere, before circling the BH at $r=r_c$. Moreover, at $r=r_c$ one also has
\begin{equation}\label{VeffSecond}
V''_{\textrm{eff},c}=-2\eta_c^2 \frac{L^2}{r_c^4},
\end{equation}
where
\begin{equation}
\eta_c = \frac{1}{2}\sqrt{4f_c-2r_c^2f''_c}.
\end{equation}
The study of instability associated with the circular orbits of massless particles on the photon sphere follows \cite{CardosoMirandaBertiWitekZanchin2009}. Indeed, writing down a second order Taylor series expansion of $V_{\textrm{eff}}$ near the photon sphere at $r=r_c$ gives
\begin{equation}\label{Taylor_EffPotential}
V_{\textrm{eff}}(r) \simeq V_{\textrm{eff},c} + \frac{1}{2}V''_{\textrm{eff},c} (r-r_c)^2,
\end{equation}
such as (\ref{EqMotion}) becomes
\begin{equation}
\dot{r}^2+V_{\textrm{eff},c} +\frac{1}{2}V''_{\textrm{eff},c} (r-r_c)^2 = E^2.
\end{equation}
If the turning point is close enough to $r=r_c$, one can consider that $E^2 \approx V_{\textrm{eff},c}$. The equation of motion then simplifies to
\begin{equation}
\dot{r}^2 +\frac{1}{2}V''_{\textrm{eff},c} (r-r_c)^2 = 0,
\end{equation}
which can be written, in $(r,t)$ coordinates, as
\begin{equation}
\left(\frac{dr}{dt}\right)^2 + \frac{V''_{\textrm{eff},c}}{2 \dot{t}^2} (r-r_c)^2 = 0.
\end{equation}
Close to $r=r_c$, one has $\dot{t}=dt/d\lambda \approx E/f_c$, and the equation of motion in the vicinity of the photon sphere simply reads
\begin{equation}
\left(\frac{dr}{dt}\right)^2 - \Lambda_c^2 (r-r_c)^2 = 0,
\end{equation}
where 
\begin{equation}\label{Lyapunov}
|\Lambda_c|=\eta_c \frac{\sqrt{f_c}}{r_c}
\end{equation}
is the Lyapunov exponent associated with the unstable circular motion of a free-falling massless particle near the BH photon sphere. 

\subsection{\label{subsec:HorizonRegion}The near horizon region: the highly damped QNM}
In this short section, we consider the well-known case of a purely radial motion in the line element $(\ref{metric_BH})$
\begin{equation}\label{metric_BH_radial}
ds^2=f(r)dt^2-\frac{dr^2}{f(r)}.
\end{equation}
In the near-horizon region, one can approximate $f(r)$ by $f(r)=f'_h(r-r_h)$ at first order, with $f'_h=(df/dr)(r=r_h)$. In a coordinate systems for which
\begin{equation}\label{rhoRindler}
d\rho=\frac{dr}{\sqrt{f'_h(r-r_h)}}
\end{equation}
and introducing the surface gravity $\kappa=(1/2)f'_h$, the Rindler approximation of the metric $(\ref{metric_BH_radial})$ in the near horizon region follows easily
\begin{equation}\label{RindlerSimple}
ds^2=\kappa^2\rho^2dt^2-d\rho^2
\end{equation}
The form $(\ref{RindlerMetric})$ is obviously recovered from which an equation similar to $(\ref{MasslessMotion2})$ is derived. It should be noted that the equation of motion of a free-falling massless particle can also be obtained from $(\ref{EqMotion})$, with $L=0$ because $\varphi$ is zero for purely radial motions towards the BH, which reads
\begin{equation}
\dot{r}^2=E^2.
\end{equation}
With $(\ref{EL})$, one deduces
\begin{equation}
\left(\frac{dr}{dt}\right)^2 - f(r)^2=0.
\end{equation}
In the near horizon region, $f(r)=f'_h(r-r_h)$ at first order in $r$, and with the change of variable $(\ref{rhoRindler})$ one obviously recovers $(\ref{MasslessMotion2})$ in the form
\begin{equation}
\left(\frac{d\rho}{dt}\right)^2-\kappa^2\rho^2=0.
\end{equation}

All the results of section \ref{subsec:ScatteringRindler} can now be applied to the near horizon case with the $SL(2,\mathbb{R})$ algebra approach. In particular, the highly damped QNF are given by a relation like $(\ref{Rindler_QNF})$
\begin{equation}
\omega_{n}^{(\pm)} \approx \mp i \kappa n,
\end{equation}
for which the imaginary part is equally spaced as expected. It should be noted that this approach does not give the real part of the QNF that appears in $(\ref{HighlyDampedQNF})$, which is related to the behavior of the BH metric near the singularity $r\to 0$.\\

\noindent Of course, although almost everything has already been said in the literature about the Hawking-Unruh effect in the near horizon region, this $SL(2,\mathbb{R})$ algebraic approach is worth mentioning, because it gives easily the equally spaced overtone level of the imaginary part of the highly damped QNF. Moreover, an interpretation that is less known is that the BH horizon appears also in this context as an unstable point, as for $(\ref{MasslessMotion2})$, here for purely radial null geodesics, in the phase space of an accelerated observer $(d\rho/dt,\rho)$. In this context, the surface gravity plays a role of a Lyapunov exponent, i.e. an inverse characteristic time, that characterizes the (in)stability of the BH horizon for radial trajectories of massless particles seen from an accelerated observer $(t,\rho)$ (see \cite{DaluiMajhi2020} and references therein for a very exhaustive study of this interpretation).

\subsection{\label{subsec:PhotonSphereRegion}The near photon sphere region: the weakly damped QNM}

\subsubsection{Geodesics near the photon sphere region}
In this section, we reproduce the outline of the proof in \cite{Raffaelli2022} according to which, in the ultrarelativistic limit, the near photon sphere limit of the metric $(\ref{metric_BH})$ is also a Rindler metric. Then, from the above, all the results of section \ref{sec:Rindler} could also be easily applied.\\ 

We start by considering the motion of a free-falling test particle of mass $m$ following the geodesic line element $(\ref{metric_BH})$. For a massive test particle to get very close to the photon sphere, we will need to consider the ultrarelativistic limit of its motion.\\

Without loss of generality, let us focus again on the equatorial plane $\theta=\pi/2$ of the metric $(\ref{metric_BH})$
\begin{equation}\label{geodesics_equatorial_plane}
ds^2=-f(r)dt^2+\frac{dr^2}{f(r)}+r^2d\varphi^{2}.
\end{equation}
In the massive case, the integrals of motion read
\begin{equation}\label{EL_massive}
f(r)\left(\frac{dt}{d\tau}\right)=\frac{E}{m}\quad ; \quad r^2\left(\frac{d\varphi}{d\tau}\right)=\frac{L}{m}
\end{equation}
where $\tau$ is the masssive particle proper time.
Using (\ref{EL_massive}), the equation of motion for the test particle reads
\begin{equation}\label{EqMotionMassive}
m^2\left(\frac{dr}{d\tau}\right)^2 + U_\textrm{eff}(r) = E^2,
\end{equation}
where 
\begin{equation}
U_\textrm{eff}(r)=f(r)\left[\frac{L^2}{r^2}+m^2\right].
\end{equation}
The effective potential $U_\textrm{eff}(r)$ has extrema located at $r=r_i$ such that
\begin{equation}\label{r0}
f'(r_i)-\frac{2}{r_i^2}f(r_i)+\frac{m^2}{L^2}f'(r_i)=0.
\end{equation}
We simplify the study by assuming that the particle angular momentum $L$ value is such that $U_\textrm{eff}(r)$ admits a local maximum. Let us call $r_0(L)$ the location of this local maximum. In the limit $L \gg 1$, $(\ref{r0})$ tends to $(\ref{rc})$, the local maximum coincides with the location of the photon sphere, i.e. $r_0(L)$ tends to $r_c$ and $U_\textrm{eff}(r_0(L))$ tends to $V_\textrm{eff,c}$.\\

In other words, a test particle (with energy $E$ and angular momentum $L$) coming from infinity, gets very close to the photon sphere if at least
\begin{subequations}\label{EL_conditions}
\begin{eqnarray}
&m^2& < E^2 \approx U_\textrm{eff}(r_0(L)),\label{E_condition}\\
&L& \gg 1 \quad \textrm{with} \quad L/E \quad \textrm{finite} \label{L_condition}.
\end{eqnarray}
\end{subequations}
The condition (\ref{E_condition}) implies that the turning point of the particle motion is located in the vicinity of the maximum of $U_\textrm{eff}(r)$. The second condition (\ref{L_condition}) implies that the location of this maximum tends to the location of the photon sphere. If the conditions (\ref{EL_conditions}) are both satisfied, then one has $E^2 \approx U_\textrm{eff}(r_0(L)) \approx V_\textrm{eff,c}$, i.e. $L/E \approx r_c/\sqrt{f_c}$, which corresponds to the limit of an ultrarelativistic test particle. Then, in the ultrarelativistic limit, the test particle has an impact parameter $b$ which tends, from above, to the critical impact parameter $b_c$ associated with massless particles motion, and the particle coming from infinity will get close to the photon sphere before moving away, back to infinity.\\

Let us now focus on the near-photon sphere limit of $(\ref{geodesics_equatorial_plane})$ to describe the motion of the test particle in the ultrarelativistic limit. We first use the constants of motions $(\ref{EL_massive})$ for geodesic motions \cite{Rindler2006}, to restrict ourselves to an equivalent effective geodesic line element in the $(t,r)$-plane. From $(\ref{EL_massive})$, one can write
\begin{equation}
r^2 d\varphi^2=f(r)^2\frac{L^2}{E^2 r^2}dt^2.
\end{equation} 
This allows to transform $(\ref{geodesics_equatorial_plane})$ into a line element, that would give the same radial equation of motion $(\ref{EqMotionMassive})$,
\begin{equation}\label{metric_BH2}
ds^2=f(r)\left(-1+\frac{V_\textrm{eff}(r)}{E^2}\right)dt^2+\frac{dr^2}{f(r)},
\end{equation}
where $V_\textrm{eff}(r)$ is still, from a pure formal point of view, defined by expression $(\ref{EffPotential})$ but with $L$ being now the angular momentum of the massive test particle. It should be noted that the line elements (\ref{metric_BH2}) and (\ref{geodesics_equatorial_plane}) have the same magnitude for any geodesic motion, i.e. for any given $E$ and $L$. Moreover, let us emphasize that (\ref{metric_BH2}) does not describe a purely radial motion in the BH background (in such case $\varphi$ would have been constant, i.e. $L=0$, and the photon sphere would have had no effect on the test particle motion), but rather describes the effective non-radial geodesic motion of a test particle in the $(t,r)$-plane of a static and spherically symmetric BH background, taking into account explicitly the effect of the centrifugal potential barrier in its time component. Let us recall that we will not consider the case where the particle gets trapped into the BH, i.e. here one has $b \gtrsim b_c$.\\ 

The near-photon sphere limit of $(\ref{metric_BH2})$ requires the conditions $(\ref{EL_conditions})$ to be satisfied, i.e. $E^2 \approx V_\textrm{eff,c}$, and is obtained from the lowest order Taylor series expansion around $r=r_c$ which does not cancel the time component of $(\ref{metric_BH2})$. Using $(\ref{Taylor_EffPotential})$, the effective line element $(\ref{metric_BH2})$ then becomes in this limit
\begin{equation}
ds^2 \simeq -\frac{V''_\textrm{eff,c}}{2E^2}f_c(r-r_c)^2dt^2+\frac{dr^2}{f_c}.
\end{equation}
Now, from $(\ref{VeffSecond})$ and $(\ref{Lyapunov})$, and introducing the variable
\begin{equation}
\rho = \frac{r-r_c}{\sqrt{f_c}} \Leftrightarrow d\rho = \frac{dr}{\sqrt{f_c}},
\end{equation}
one finally obtains a Rindler form of the effective line element near the photon sphere, which acts in this setting as an effective Rindler horizon
\begin{equation}\label{Rindler_metric}
ds^2 \simeq - \Lambda_c^2 \rho^2 dt^2 + d\rho^2,
\end{equation}
where we have used $L^2/E^2 \approx r_c^2/f_c$ because $E^2 \approx V_\textrm{eff,c}$.\\
The Lyapunov exponent $|\Lambda_c|$ associated with the massless particle motions around the photon sphere plays the role of a constant proper acceleration in the near-photon sphere limit of the line element (\ref{metric_BH2}), describing the test particle effective geodesic motion in the $(t,r)$-plane, a role analogue to the role played by the surface gravity in the near-horizon limit.

\subsubsection{Massless scalar field in the BH metric}
In this section, we look at the Klein-Gordon equation for a massless scalar field $\Phi$ in a spacetime metric $(\ref{metric_BH})$. After separation of variables, assuming a harmonic time dependence ($e^{-i\omega t}$) for $\Phi$ and the introduction of the radial partial wave functions $\Phi_{\ell\omega}(r)$ with $\ell=0,1,2,\dots$, the Klein-Gordon equation gives the well-known Regge-Wheeler equation

\begin{equation}\label{ReggeWheeler}
\frac{d^2 \Phi_{\ell\omega}}{d r_\ast^2} + \left[\omega^2 - V_\ell(r_\ast)\right] \Phi_{\ell\omega} = 0.
\end{equation}
where $r_\ast = r_\ast(r)$ is the tortoise coordinate, introduced previously, and $V_{\ell}(r)$ is the Regge-Wheeler potential defined as
\begin{equation}\label{RWPotential}
V_{\ell}(r)=f(r)\left[\frac{\ell(\ell+1)}{r^2}+\frac{1}{r}f'(r)\right].
\end{equation}

\noindent It should be noted that, for every $\ell \in \mathbb{N}$, $V_{\ell}(r)$ admits a local maximum at $r=r_{0}(\ell)$ which is close to the photon sphere located at $r=r_c$. Moreover, in the limit $\ell \gg 1$, $r_0(\ell) \approx r_c$ and $V_{\ell}(r) \approx V_{\textrm{eff}}(r)$. Using the tortoise coordinate, we will denote $(r_\ast)_{0,\ell} = r_\ast(r_0(\ell))$ the location of the maximum of $V_{\ell}(r)$.\\

\noindent Following \cite{SchutzWill1985}, let us consider $V_\ell(r_\ast)$ around the location of its local extremum at $(r_\ast)_{0,\ell}$, i.e. around a location as close to the photon sphere as $\ell \gg 1$. A second order Taylor series expansion gives
\begin{equation}\label{VTaylorSeries}
V_\ell(r_\ast) \approx V_0(\ell) + \frac{1}{2}V^{(2)}_0(\ell) \left(r_\ast - (r_\ast)_{0,\ell} \right)^2,
\end{equation}
where we have used the notation
\begin{equation}
V_0^{(2)}(\ell)=\left(\frac{d^2 V_{\ell}(r_\ast)}{dr_\ast^2}\right)_{(r_\ast)_{0,\ell}}.
\end{equation}
Introducing $x$, $h(\ell,\omega)$ and $\alpha(\ell)$ such that
\begin{eqnarray}\label{VariablesPhotonSphereLimit}
x &=& \left(r_\ast - (r_\ast)_{0,\ell} \right),\nonumber\\ 
h(\ell,\omega) &=&\omega^2-V_0(\ell),\nonumber\\
\alpha(\ell) &=& \sqrt{-2V_0^{(2)}(\ell)},
\end{eqnarray}
equation (\ref{ReggeWheeler}) reads

\begin{equation}\label{EigenvaluePb}
H\tilde{\Phi}_{\ell\omega}(x) = h(\ell,\omega)\tilde{\Phi}_{\ell\omega}(x)
\end{equation}
where $\tilde{\Phi}_{\ell\omega}(x)=\Phi_{\ell\omega}(r_\ast(x))$ and 
\begin{equation}
H=-\frac{d^2}{dx^2} - \frac{\alpha(\ell)^2}{4}x^2
\end{equation}
is the time-independent ``Hamiltonian'' (here with dimension $L^{-2}$) governing the massless scalar field dynamics in the near-photon sphere limit. In the coordinate $x$, $H$ is in the form $(\ref{Hamiltonian})$, and then all the eigenvalue problem can be solved via algrebraic $SO(2,1)\sim SL(2,\mathbb{R})$ method.\\ 


For the weakly damped BH QNF (i.e. $\ell \gg 1$ and $\ell\gg n$), it should be noted that one can make the substitution $\ell(\ell+1)=(\ell+1/2)^2-1/4$ in $(\ref{RWPotential})$ and look at $(\ell+1/2)$ (instead of $\ell$) as the quantity in which one will compute a Taylor series expansion of the Regge-Wheeler potential and its second derivative at $(r_{\ast})_{0,\ell}$, in the limit $(\ell+1/2)\gg 1$. At lowest order in $\ell+1/2$, one has
\begin{eqnarray}
\alpha(\ell) &\approx & 2\eta_c \frac{f_c}{r_c^2} \left(\ell+\frac{1}{2}\right),\\
V_{0}(\ell) &\approx & \frac{f_c}{r_c^2}\left(\ell+\frac{1}{2}\right)^2. 
\end{eqnarray}
Equation $(\ref{HEigenvalues})$ then reads 
\begin{equation}
\omega^2=V_{0}(\ell) \pm i \alpha(\ell)\left(n+\frac{1}{2}\right),
\end{equation}
which gives, in the weakly damped regime ($(\ell+1/2)\gg 1$ and $\ell \gg n$), a equally spaced imaginary part of the BH QNF
\begin{equation}\label{QNF}
\textrm{Im}\left(\omega_{\ell n}^{(\pm)}\right)\approx -\frac{i}{2}\frac{\alpha(\ell)}{\sqrt{V_0(\ell)}}\left(n+\frac{1}{2}\right)=-i|\Lambda_c|\left(n+\frac{1}{2}\right),
\end{equation}
with $|\Lambda_c|=\eta_c \sqrt{f_c}/r_c=$ is the Lyapunov exponent that characterizes the instability of null motions on the photon sphere.

\section{\label{sec:Rks}Remarks on the amplitude of the overtone spacing}
A relevant fact is that the amplitude of the overtone level spacing is related to a Lyapunov exponent that characterizes the (in)stability of the top of the underlying IHO potential barrier. Rindler spacetime is an example of such behavior: the surface gravity in $(\ref{RindlerSimple})$ is understood in this framework as a Lyapunov exponent, and reciprocally, the Lyapunov exponent in $(\ref{Rindler_metric})$ can be interpreted as a locally constant acceleration for a certain family of observers. It should be noted that in both cases, the Lyapunov exponent characterizes motions of massless particles that seem to play a fundamental role in this description. Of course, as long as one has a Rindler approximation of a BH metric, the thermal aspects follow directly (see \cite{DaluiMajhi2020,Raffaelli2022} for more details). As a Lyapunov exponent is interpreted as the inverse of a characteristic time, this allows to expect a fundamental link between ``time-thermality-instability''.\\

\noindent Another remark is about the real part of the QNF in the highly damped regimes. This approach does not allow to access such a real part, because it has been shown to be related to the behavior of the BH metric near its singularity $r\to 0$, which is not considered here.


\section{\label{sec:Cl}Conclusion}
In this paper, we propose a simple way to understand the equally spaced overtone level of BH QNF both in the highly and weakly damped regimes. To do so, we have shown that the resonant scattering problem in Rindler spacetime is deeply linked to the resonant scattering problem by an IHO potential, which in turn can be completely described through the $SL(2,\mathbb{R})$ algebra. The $SL(2,\mathbb{R})$ algebra allows to show that the spectrum of the ``Hamiltonian'' is discrete, purely imaginary and indexed by a non negative integer. This integer turns out to be simply the overtone of the QNF, and is shown to be equally spaced. A general claim could follow: in any BH geometry, at last for static and spherically symmetric BH, if there exist some regions where the BH metric can be locally approximated as a Rindler spacetime, in some set of coordinates, then the underlying algebra should be $SL(2,\mathbb{R})$, and the imaginary part of the BH QNF, be it in the highly damped or weakly damped regimes, will be indexed by a non negative integer, i.e. the overtone number, which will be equally spaced. A very interesting fact is that we explicitly obtained this result for both highly and weakly damped QNM families, by looking at different regions of the BH spacetime. This work can be easily extended to higher dimensional spacetime. A future look at the Kerr BH is of interest.

\begin{acknowledgments}
The author would like to thank J.P. Provost and J.L. Jaramillo for stimulating discussions. The IMB receives support from the EIPHI Graduate School (contract ANR-17-EURE-0002).
\end{acknowledgments}





\bibliography{BR_QNMovertone_v1bib}

\end{document}